\documentclass[11pt]{article}

\usepackage[margin=1in]{geometry}
\usepackage{amsmath,amssymb,amsthm}
\usepackage{mathtools}
\usepackage{booktabs}
\usepackage{microtype}
\usepackage[hidelinks]{hyperref}
\usepackage[round]{natbib}
\usepackage{enumitem}
\theoremstyle{plain}
\newtheorem{theorem}{Theorem}

\theoremstyle{definition}
\newtheorem{axiom}{Axiom}

\theoremstyle{remark}

\newcommand{\E}{\mathbb{E}}
\newcommand{\KL}[2]{D\!\left(#1\,\|\,#2\right)}
\newcommand{\I}{I}
\DeclareMathOperator{\Cov}{Cov}

\title{\bf A Mathematical Theory of Value\\[4pt]
{\large\normalfont a synthesis on goal-directed agency under resource constraints}}
\author{Cheng Qian}
\date{2026}

\begin{document}
\maketitle

\begin{abstract}
We propose that \emph{value} --- the quantity that goal-directed agents create, destroy, and exchange --- is a
lawful structural quantity, in the same category as information, once stripped of its semantic clothing
(morality, price, psychology). Following the method of \citet{shannon1948}, we make one ruthless abstraction:
value is the rate at which an agent converts a physical resource into goal-progress, relative to a frame fixed
by the agent's goal. A scale-invariance axiom forces a logarithmic measure of value, $V=\sum_i k_i\ln e_i$, via a Cauchy
functional equation; the compounding dynamics of a reinvested resource force the same form independently via
the ergodicity argument of \citet{peters2019}. The two routes are kin rather than independent --- the
scale-invariance axiom is the static shadow of multiplicative compounding --- so their agreement is a
consistency check on the form, not an independent over-determination. From the compounding dynamics we also derive a \emph{coding theorem of value}: the rate at which an agent can create value through a perception channel $Y$ of
the world $X$ is bounded by the mutual information, $\Delta G \le \I(X;Y)$, achieved by Bayes-proportional
allocation; and realized value decomposes exactly as available potential minus dissipation,
$G=\KL{q}{r}-\KL{q}{p}$, identifying misalignment with measurable waste. For populations we show value is
frame-relative while \emph{price} is frame-independent, that a fleet which pools its resource and fuses its
perception is one agent and so inherits the capacity ceiling, $G_{\mathrm{fleet}}\le \I(X;Y_{1:m})\le H(X)$ (a
corollary; an earlier sum-form claim was wrong and is corrected in v5), and that the fleet's operating point is a Kelly portfolio
over agents selected by an emergent price. A dynamical layer gives the equations of motion and an
is/ought asymmetry --- beliefs have a target the world supplies, goals do not --- from which alignment emerges
as a control-stability condition with a closed-form residual misalignment. We then test the single-frame laws on
live language models in a \emph{pre-registered} scale-up across three task domains and a ten-model, five-family
ladder ($0.5$B--$8$B): perception mutual information tracks realized \emph{capability} rather than parameter count
(Spearman $\rho = 0.977$ pooled over 30 model$\times$domain points), out-of-sample $\Delta G$ tracks $\I(X;Y)$
(slope CI excludes $0$), and over-confidence is measurable dissipation in every domain --- the two single-frame
laws generalize. A further pre-registered test shows the out-of-sample bridge $\Delta G\sim\I(X;Y)$ is
\emph{shape-invariant}: pooled across four qualitatively different task shapes --- classification, reasoning
(GSM8K), sequential decision, and code (MBPP), all with discrete gold so $\I$ is computed, not estimated ($n=42$)
--- the slope is $0.953$, statistically indistinguishable from the classification-only value, promoting the
bridge from a demonstration toward a law. A fleet-pricing experiment is reported scoped and primary-metric-first: value-pricing recovers
cost-aware routing from first principles --- it ties good hand-tuned routing under a token budget, beats a
cost-blind router under a compute budget, and does not outperform a cost-aware engineer; its contribution is
principled measurement, not outperformance. The paper's stated continuation gate has since been run
(pre-registered, on a frontier-model population): the coupled capacity-region prediction --- the growth-gap law,
coalition submodularity with a designed XOR synergy control, the joint ceiling, and Kelly selection --- is
\emph{confirmed within its frozen bands} on real agents, the first real-agent confirmation of a prediction no
component theory makes separately; the mean-field residual-scaling law $\|Vg\|/\gamma$, by contrast, found
\emph{no domain} --- capable populations hold no goal dispersion ($V\to 0$) --- and is retired to its
mathematical scope. The laws hold in the smooth, concave (diminishing-returns) regime;
threshold, satiation, and risk-seeking goals lie outside it. None of the underlying mechanisms is individually
new --- the single-agent core is generalized Kelly, the is/ought asymmetry's value-side is \citet{armstrong2018occam},
and the alignment-stability algebra is classical control; the contribution is their unification under one
substrate-grounded quantity and the governance mapping (incentive design over oversight) that follows.
\end{abstract}

\tableofcontents

\section{Introduction}\label{sec:intro}

Information theory began with an abstraction that looked like a loss. \citet{shannon1948} threw away
\emph{meaning} and kept only the reduction of uncertainty; from that single move came a measure ($-\log p$), a
unit (the bit), and hard limits (channel capacity, the coding theorems). The abstraction was productive
precisely because it was ruthless: by refusing to model what a message \emph{meant}, Shannon could state exactly
what any communication system can and cannot do.

We attempt the analogous move for \emph{value}. The everyday notion of value is a tangle of morality, market
price, and human preference. We throw all three away and keep one structural residue: \emph{value is the rate at
which a goal-directed agent converts a scarce physical resource into progress toward its goal.} This is
frame-relative --- it is defined only relative to a goal, the way information is defined only relative to a prior
and energy only relative to a reference frame --- but relativity does not disqualify a quantity from being
fundamental. It forces the theory to make the frame explicit. A theory of value is to agency what relativity is
to motion: the quantity is frame-dependent; the laws relating frames are universal.

The motivation is not only foundational. If, as increasingly seems likely, near-future populations of
artificial agents are each driven by an explicit objective, then a theory of value is not philosophy --- it is
the control theory for agent populations. To route, align, and govern many separated agents we need value to be
a \emph{measured} quantity with conservation laws and limits, not a verdict. This paper builds that quantity and
tests it on real agents.

\paragraph{Contributions and honesty.} We claim none of the underlying mechanisms as individually novel. The
logarithm is Bernoulli's and Kelly's; the growth-rate-as-information-rate is \citeauthor{kelly1956}'s; the
price layer is Arrow--Debreu's; the information geometry is \v{C}encov's and Amari's. In particular --- to meet
the concession here rather than leave it for a referee --- our capacity theorem (Theorem~\ref{thm:capacity}) \emph{is}
generalized Kelly; what we add is not the betting result but everything built upon it: the free-energy substrate
that supplies a unit, the cross-frame price, and the dynamics and alignment layers. Our contribution is
fourfold and we state it plainly (\S\ref{sec:related}): (i) the \emph{unification} of these accounts under one
substrate-grounded quantity; (ii) the \emph{capacity theorem} and the \emph{pooled-fleet ceiling} that the
unification makes visible (both corollaries of Kelly--Cover, claimed as such); (iii) the \emph{is/ought
asymmetry} as a structural property of agent dynamics; and (iv) an \emph{empirical test} on live models. Throughout, each result is followed by the assumptions that make it bite,
in Shannon's spirit of naming the model.

\bigskip
\noindent\textbf{Plan.} Part~\ref{part:statics} (statics) derives the measure, the capacity theorem, and the
Second Law for a single agent. Part~\ref{part:multi} (multi-agent) develops cross-frame price and the fleet
capacity region. Part~\ref{part:dynamics} (dynamics) gives the equations of motion and alignment-as-stability.
Part~\ref{part:experiments} (evidence) reports the synthetic and real-agent tests. Appendices collect the two
proofs.

\part{Statics: the measure, the limit, the Second Law}\label{part:statics}

\section{The measure: a logarithmic law of value}\label{sec:measure}

\paragraph{Setup.} An agent holds a divisible resource of total size $E>0$ (free energy, compute, budget,
attention --- any fungible substrate of action) and allocates it across $K$ goal-relevant channels in amounts
$e=(e_1,\dots,e_K)$, $\sum_i e_i=E$. Each channel $i$ carries a \emph{goal weight} $k_i\ge 0$ measuring how much
progress toward the goal a unit of effective work on $i$ yields. Let $V(e)$ be the value the allocation
realizes. Two independent structural arguments force the same logarithm; their agreement is the sign of a real law.

\paragraph{Confirmation I --- static: axioms uniquely fix the functional form.}
The purely static question --- what shape must $V$ take to be a consistent value measure? --- is answered
uniquely by three axioms via Cauchy's theorem (Appendix~\ref{app:log}).

\begin{axiom}[Diminishing returns]\label{ax:concave}
$V$ is increasing and strictly concave in each $e_i$: the first unit of resource on a channel is worth more than
the thousandth.
\end{axiom}
\begin{axiom}[Additivity across independent channels]\label{ax:additive}
For channels whose contributions do not interact, $V(e)=\sum_i v(e_i;k_i)$.
\end{axiom}
\begin{axiom}[Scale invariance]\label{ax:scale}
Rescaling the resource unit ($e_i\mapsto \lambda e_i$) changes total value only by an additive constant
independent of the allocation: $v(\lambda e_i;k_i)=v(e_i;k_i)+c(\lambda,k_i)$.
\end{axiom}

Axiom~\ref{ax:scale} is the load-bearing one for this route: it encodes the absence of a privileged
resource-unit (value has no absolute zero of scale, just as information has no privileged base of logarithm).
It is a Cauchy functional equation in disguise.

\paragraph{Confirmation II --- dynamic: compounding forces the same form.}
The dynamic question --- what is the correct measure of long-run gain when value compounds? --- is answered
by the ergodicity argument of \citet{peters2019} and, independently, \citet{kelly1956}: the time-average
growth of a multiplicatively-reinvested resource is the expected log of its multiplier. Section~\ref{sec:capacity}
develops this rigorously via the repeated game and horse-race structure, showing that the optimal long-run
allocation realizes $V^\star=\sum_i k_i\ln e_i$ --- the same logarithmic form, derived from dynamics with
\emph{no} scale assumption. The two confirmations address different questions: Axioms~\ref{ax:concave}--\ref{ax:scale}
fix the static functional form; the compounding route shows the dynamic optimum lands on the same shape.

\begin{theorem}[Logarithmic Value Law]\label{thm:log}
Under Axioms~\textup{\ref{ax:concave}--\ref{ax:scale}} with mild regularity, the value measure is, up to choice
of unit,
\begin{equation}
\boxed{\;V(e)\;=\;\sum_{i=1}^{K} k_i \ln e_i\;}
\end{equation}
and the value-maximizing allocation of a fixed budget $E$ is proportional to the goal weights,
$e_i^\star = E\,\hat k_i$ with $\hat k_i=k_i/\sum_j k_j$, giving
\begin{equation}
V^\star \;=\; K\ln E \;-\; K\, H(\hat k)\,,\qquad H(\hat k)=-\textstyle\sum_i \hat k_i\ln\hat k_i .
\end{equation}
\end{theorem}
\noindent Two features deserve note. First, Shannon's entropy $H$ reappears \emph{unbidden} as a penalty: a goal
spread thinly over many channels (high $H(\hat k)$) realizes less value than a focused one ---
\emph{focus is worth $K\,H(\hat k)$ nats of value}. Second, the logarithm arrives twice: the static
Axioms~\ref{ax:concave}--\ref{ax:scale} force it via the Cauchy equation (Appendix~\ref{app:log}), and
the dynamic compounding argument recovers it from \S\ref{sec:capacity}. We are explicit that these routes are
kin rather than independent --- the scale-invariance axiom is the static shadow of multiplicative compounding
(a parallel additive axiom would force a linear $v$ by the identical Cauchy argument) --- so their agreement
is a consistency check on the functional form, not an independent over-determination.

\section{The capacity theorem: value is bounded by information}\label{sec:capacity}

\paragraph{The repeated game.} Now let value compound: value realized this round becomes resource next round (an
agent that achieves its goal earns more budget, trust, or compute). Each round $t$ the world settles into a
state $s\sim q$; the agent splits its budget into fractions $b=(b_1,\dots,b_n)$, $b_i\ge0$, $\sum_i b_i=1$; and
channel $i$ returns a multiplier $o_i(s)\ge 0$ per unit committed. The budget updates multiplicatively,
$E_{t+1}=E_t\sum_i b_i\,o_i(s_t)$, so $\log E_T=\log E_0+\sum_t \log\!\big(\sum_i b_i o_i(s_t)\big)$ and the
long-run \emph{value growth rate} is
\begin{equation}\label{eq:G}
G(b)\;=\;\E_{s\sim q}\!\Big[\log\textstyle\sum_i b_i\,o_i(s)\Big].
\end{equation}
That the time-average growth of a multiplicative process is the \emph{expected log} of its multiplier is the
ergodicity argument of \citet{peters2019}; it forces the logarithm again, now from dynamics rather than
psychology. $G$ is concave in $b$, so maximizing it is a well-posed convex program.

\paragraph{The horse race.} Take the canonical structure in which states are the channels, committing to channel
$i$ pays only in state $i$, and returns are quoted against a \emph{reference belief} $r$ over states,
$o_i=1/r_i$. Then $G(b)=\sum_s q_s\log(b_s/r_s)$, maximized (Gibbs) by betting one's model,
$b_s=p_s$. With a correct model $p=q$,
\begin{equation}\label{eq:Gstar}
G^\star=\sum_s q_s\log\frac{q_s}{r_s}=\KL{q}{r}.
\end{equation}
\emph{Maximal value growth equals the divergence of the agent's correct model from the baseline.} An agent whose
model is the baseline ($p=r$) grows value at rate zero: value creation is informational edge over the prevailing
expectation, nothing more.

\paragraph{Side information.} Equip the agent with a perception channel: a signal $Y$ correlated with the
world-state $X$, observed before allocating. Optimal play bets the posterior $b_s=p(s\mid y)$, with growth
$G_Y=\sum_{x,y}p(x,y)\log\frac{p(x\mid y)}{r(x)}$. The gain over acting on the prior alone is exactly mutual
information:
\begin{theorem}[Coding Theorem of Value]\label{thm:capacity}
The incremental rate at which an agent can compound value by perceiving the world through a channel $Y$ is at
most the mutual information between world-state and perception,
\begin{equation}
\boxed{\;\Delta G \;=\; G_Y-G_0 \;=\; \I(X;Y)\;,}
\end{equation}
and this bound is achieved by Bayes-proportional allocation. \textup{(Proof: Appendix~\ref{app:capacity}.)}
\end{theorem}
\noindent This has the full structure of a Shannon coding theorem. \emph{Achievability:} betting the posterior
attains $\Delta G=\I(X;Y)$. \emph{Converse:} any $b\ne p(\cdot\mid y)$ loses exactly $\KL{p(\cdot\mid y)}{b}$ per
round (Gibbs), so $\I(X;Y)$ is an upper bound. An agent cannot create value faster than it can perceive the
world; \emph{value-throughput is bounded by information-throughput.} Perception capacity is the hard ceiling on
value-generation rate --- the value analog of channel capacity.

\section{The Second Law of Value}\label{sec:secondlaw}

Drop the assumption that the agent's model is correct. Let $q$ be reality, $p$ the agent's model, $r$ the
baseline; the agent bets $b=p$ but the world is drawn from $q$. Then
\begin{equation}\label{eq:secondlaw}
\boxed{\;G_{\mathrm{actual}}=\sum_s q_s\log\frac{p_s}{r_s}
=\underbrace{\KL{q}{r}}_{\text{available potential}}-\underbrace{\KL{q}{p}}_{\text{dissipation (model error)}}\;.}
\end{equation}
Realized value is available potential minus dissipation from misalignment, every term in nats.

\emph{The name is a structural analogy, stated with explicit scope.} It holds exactly where thermodynamic
entropy does: the dissipation $\KL{q}{p}\ge 0$ is non-negative, \emph{destroyed} rather than created, and
monotonically reduced as the agent learns ($p\to q$). It \emph{breaks} where entropy does not: value here is
goal- and frame-relative --- fixed by the reference $r$ and the goal-weights $k$ --- not an observer-independent
quantity, so this is a Second Law \emph{of value}, not a corollary of statistical mechanics. We keep the name
because the formal structure --- a non-negative, non-creatable dissipation that learning minimizes --- is
genuinely shared, and bound it because the substrate is not.

Three consequences. (i) \emph{Misalignment is measurable dissipation}: confident error ($\KL{q}{p}>\KL{q}{r}$) drives
growth negative --- a wrong, certain agent actively destroys value, while calibrated humility is
value-preserving. (ii) \emph{Learning is value-recovery}: $\KL{q}{p}$ is exactly the cross-entropy excess that
predictive training minimizes, so the machine-learning objective \emph{is} the minimization of value
dissipation. (iii) \emph{Alignment is a value law}: reducing $\KL{q}{p}$ is mechanically identical to reducing
wasted value. We test \eqref{eq:secondlaw} on live models in \S\ref{sec:real} (R2), where the weakest model's
over-confidence indeed drives realized growth negative.

\part{Multi-agent: price and the fleet}\label{part:multi}

\section{Cross-frame value and the frame-independence of price}\label{sec:crossframe}

A single agent's value lives in its own frame $(q,p,r,k)$. Two agents with different goals have values that are
\emph{not} cardinally comparable --- the interpersonal-comparison impossibility of \citet{arrow1951}, which we
concede rather than circumvent. Yet they can still coordinate, exactly as economies coordinate non-comparable
utilities: through a \emph{price} on the shared resource.

\paragraph{Shadow price.} For an agent with goal mass $K=\sum_i k_i$ and budget $E$, the marginal value of
resource (the Lagrange multiplier on $\sum_i e_i=E$ in Theorem~\ref{thm:log}) is the \emph{shadow price}
$\lambda=K/E$. Two agents trade resource until their shadow prices equalize; the common $\lambda$ is a scalar
that lives in \emph{neither} agent's value frame. Hence:
\begin{quote}
\emph{Value is frame-relative; price is frame-independent.} Value is a vector in an agent's goal basis; price is
the one scalar on which separated frames agree.
\end{quote}

\paragraph{The invariant.} The transformation between two agents' value frames is a change of coordinates on the
belief simplex. The unique (up to scale) Riemannian metric invariant under such sufficient-statistic
reparametrizations is the Fisher--Rao metric \citep{cencov1982,amari2016}. It is the cross-frame invariant of
the theory: distances in belief that all agents agree on, whatever their goals. Alignment between two goals is
the cosine $\cos\theta_{ab}=\langle \hat k_a,\hat k_b\rangle$; its sign sets whether their interaction is
positive- or negative-sum.

\section{The fleet capacity region}\label{sec:fleet}

Consider $m$ agents acting on one world $X$ and drawing on one resource pool.

\paragraph{The ceiling (corrected in v5).} Each agent's growth is bounded by its own perception
(Theorem~\ref{thm:capacity}), $G_a\le \I(X;Y_a)$. An earlier version of this paper (v1--v4) additionally boxed
a \emph{sum-rate} bound, $\sum_{a\in S} G_a \le \I(X;Y_S) \le H(X)$, claimed via data-processing. \textbf{That
claim is false in the model as stated}, and external review caught it: two agents with identical perfect
channels $Y_1=Y_2=X$, each compounding its \emph{own} budget against fixed reference odds, each achieve
$G_a=H(X)$ by Theorem~\ref{thm:capacity}'s achievability, so $\sum_a G_a = 2H(X) > H(X)$. The chain rule bounds
the \emph{joint} information, not the sum of separately-compounded growth rates; for redundant channels
$\sum_a \I(X;Y_a) \ge \I(X;Y_{1:m})$ --- the wrong direction --- and the multiple-access analogy fails because
independently-bankrolled agents, unlike transmitters, share no coupling medium. What the shared world does
bound is the fleet acting as \emph{one} decision-maker: pooling the budgets and betting the fused posterior
$p(x\mid y_1,\dots,y_m)$ makes the fleet a single agent with channel $Y_{1:m}$, so Theorem~\ref{thm:capacity}
applies verbatim,
\begin{equation}\label{eq:ceiling}
\boxed{\;G_{\mathrm{fleet}} \;\le\; \I(X;Y_{1:m})\;\le\; H(X)\;}\qquad\text{(pooled bankroll, fused posterior)},
\end{equation}
a one-line corollary we do not claim as new. A sum-form constraint holds only under an explicit payout
coupling (a single shared bankroll, or parimutuel odds where agents bet against each other), stated as an
assumption for those settings. The governance corollaries survive re-scoped to the fused reading: once
$\I(X;Y_{1:m})$ saturates, redundant agents add zero to what the fleet \emph{jointly knows} — \emph{perception
diversity} lifts the fused ceiling toward $H(X)$, \emph{redundancy} adds exactly zero. We confirm the joint
(fused) version on live models in \S\ref{sec:real} (R4); the simulation suite likewise tests the joint form.

\paragraph{The operating point.} Under multiplicative dynamics the shared pool multiplies each round by
$\sum_a w_a R_a$ for resource weights $w_a$, so the fleet is a \emph{Kelly portfolio over agents}: for
imperfectly-correlated agents, spreading resource and rebalancing raises the time-average growth (the
volatility-harvesting effect of \citealp{kelly1956,cover1991}). The weights that achieve the optimum are
selected by the resource \emph{price} $\pi$: price moves resource to high-shadow-price agents until $\lambda$
equalizes. Thus
\begin{quote}
the \emph{alignment matrix} $M=[\cos\theta_{ab}]$ shapes the achievable region (a cooperation dividend where
$\cos\theta>0$, a conflict tax where $\cos\theta<0$); the \emph{price} selects the operating point on it.
\end{quote}
This separates governance into two acts: \emph{shaping} the region (alignment and perception design) and
\emph{choosing} the point on it (pricing). The pricing claim is the one we test most carefully
(\S\ref{sec:real}, R5), and where the demon's precondition --- genuine diversity --- turns out to matter.

\part{Dynamics: motion and alignment}\label{part:dynamics}

\section{The equations of motion and the is/ought asymmetry}\label{sec:dynamics}

The statics fix equilibria; the dynamics say how they are approached. Three objects evolve. \emph{Beliefs} $p$
flow toward the true law $q$ by a natural-gradient (Fisher) descent on log-loss --- the realized regret of this
flow equals the cumulative dissipation $\sum_t \KL{q}{p_t}$ of \eqref{eq:secondlaw}, so \emph{learning is
value-recovery} dynamically as well as statically. \emph{Prices} $\pi$ flow toward market-clearing by
t\^atonnement (dual ascent on the resource constraint). \emph{Goals} $k$ flow under two forces: \emph{control}
(a principal pulling $k_a$ toward a target $k^\star$) and \emph{selection} (compounding reweights resource
toward higher-growth goals).

The organizing fact is an asymmetry. Beliefs and prices each flow toward a target \emph{the world supplies}:
reality provides $q$, the resource constraint provides market-clearing. Goals have \emph{no} target the world
supplies --- reality contains no fact about what ought to be valued. So beliefs and prices are \emph{learnable}
(gradient flows toward a given target); goals are only \emph{controllable} or \emph{selectable}. This is Hume's
is/ought gap recovered as a structural property of the dynamics, and \S\ref{sec:alignment} shows it is the
mathematical shape of the alignment problem.

\section{Alignment as a stability condition}\label{sec:alignment}

Let agent $a$ have goal $k_a$ and resource share $w_a$, with fleet effective goal $\bar k=\sum_a w_a k_a$.
Control pulls each goal toward target: $\dot k_a|_{\mathrm{ctrl}}=-\gamma(k_a-k^\star)$. Selection follows the
replicator $\dot w_a=w_a(G_a-\bar G)$. Summarize the environment's reward landscape over goals to first order by
its growth gradient $g:=\nabla_k G$ --- the direction in goal-space along which deviating from $k^\star$
\emph{increases} resource capture.

\begin{theorem}[Coupled-flow alignment]\label{thm:align}
By Price's equation \citep{price1970}, the selection drift of the mean is the trait--fitness covariance; with
$G_a\approx \bar G+g^\top(k_a-\bar k)$ it equals $Vg$, where $V=\Cov_a(k_a)$ is the goal-dispersion matrix.
Hence the effective goal obeys
\begin{equation}
\dot{\bar k}=-\gamma(\bar k-k^\star)+Vg,
\end{equation}
with fixed point and residual misalignment
\begin{equation}
\boxed{\;\bar k^\star=k^\star+\gamma^{-1}Vg,\qquad \|\bar k^\star-k^\star\|=\gamma^{-1}\|Vg\|\;,}
\end{equation}
and the aligned fixed point is locally stable iff $\gamma$ exceeds the spectral abscissa (largest real part of
the eigenvalues) of $\partial(Vg)/\partial\bar k$ --- written $\gamma>\lambda_{\max}$ for the normal/symmetric
case; for non-normal Jacobians, non-normality additionally permits transient misalignment growth even when the
fixed point is asymptotically stable.
\end{theorem}
\noindent The fleet does not in general settle at the target; it settles a distance $\|Vg\|/\gamma$ away ---
goal-dispersion times reward-pull over control gain. Perfect alignment holds iff $Vg=0$: either the environment
rewards exactly the target ($g=0$), or there is no diversity to select among ($V=0$), or control is infinite.
The governance reading is sharp:
\begin{quote}
\emph{The cheap half of alignment is incentive design, not control.} Driving $g\to0$ (aligning what \emph{pays}
with what is \emph{wanted}) removes the residual for \emph{any} $\gamma$ and \emph{any} $V$, preserving the
diversity that lifts the fleet ceiling \eqref{eq:ceiling}; raising $\gamma$ (brute-force oversight) merely
opposes a drift it never removes. Spend first on $g$.
\end{quote}
This is the actionable form of the is/ought asymmetry: because goals have no world-given target, they are
governed by what pays (selection, via $g$) and what is imposed (control, via $\gamma$).

\part{Evidence}\label{part:experiments}

\section{Synthetic validation}\label{sec:sim}

We first check that each closed-form prediction is reproduced by a Monte-Carlo world built to the theory's
assumptions. All five families pass to numerical tolerance (20/20 checks): E1, $\Delta G=\I(X;Y)$ to
$\sim\!10^{-3}$ nats; E2, the Second-Law decomposition \eqref{eq:secondlaw} exactly, including value going
negative under confident error; E3, the fleet ceiling \eqref{eq:ceiling} with diversity lifting it and
redundancy adding zero; E4, a Kelly-priced fleet beating ad-hoc allocation, including a ``Shannon's demon''
built from anti-correlated agents that individually do not grow yet collectively do; E5, cumulative dissipation
equal to regret, with a drifting world flooring dissipation at a positive value (a dynamical Second Law). These
confirm the mathematics is self-consistent and correctly derived. They are, however, circular by construction:
the worlds are drawn from the same distributions the formulas assume. The decisive test is on real agents.

\section{Real agents}\label{sec:real}

\paragraph{Instantiation.} We take a frozen, held-out 100-item decision task in which each item has a
ground-truth correct action drawn from $K=7$ classes, and an agent's output \emph{is} an action. This realizes
the perception-then-act structure of Theorem~\ref{thm:capacity} directly: the world-state $X$ is the correct
action, the perception $Y_a$ is the action model $a$ chooses. We set $r=q$ (the action marginal), so a
no-signal agent grows value at rate $0$. From each model's run we form the confusion matrix
$C_a[x,y]=\#(\text{gold}=x,\text{chosen}=y)$ and compute every quantity in nats with the same primitives used in
\S\ref{sec:sim}: $\I(X;Y_a)$ from the joint; the calibrated posterior $p_a(x\mid y)=C_a[:,y]/\sum_x C_a[:,y]$;
and realized growth $\Delta G_a=\text{mean}\,\ln\!\big(p_a(x\mid y_a)/r(x)\big)$. The agents are four local
models spanning a range of realized capability --- three of one family ($1.5$B, $3$B, $7$B) and a cross-family
model (an $8$B that is, on this task, \emph{less} capable than the $7$B) --- greedy decoding. Posteriors are
calibrated on a $48$-item fit split and scored on a $52$-item holdout; in-sample, the oracle posterior with
$r=q$ makes $\Delta G=\I$ an arithmetic identity (used only to confirm the units), so the empirical content is
the capability tracking, the out-of-sample tracking, and the fleet result.

\paragraph{R1: the bridge holds, and $\I$ tracks capability, not size.} Across the four models, mutual
information increases monotonically with realized capability (tool-accuracy), and out-of-sample value-growth
increases monotonically with $\I$ (Table~\ref{tab:r1}; with only four models we report the ordered values, not
a correlation coefficient). The decisive datum is the cross-family model D: it is \emph{larger} than C (8B vs 7B) yet
\emph{less} capable ($0.86$ vs $0.96$), and its $\I$ lands accordingly below C, near B. So the claim is not
``scale buys value'' but the sharper one: value-throughput is information-throughput, and information-throughput
is set by what the model can actually perceive.

\begin{table}[h]\centering
\begin{tabular}{lccccc}
\toprule
model & params & tool-acc & $\I(X;Y)$ & $\Delta G_{\text{in}}$ & $\Delta G_{\text{hold}}$ \\
\midrule
A & 1.5B & 0.79 & 1.277 & 1.277 & 0.915 \\
B & 3B   & 0.87 & 1.561 & 1.561 & 1.242 \\
C & 7B   & 0.96 & 1.779 & 1.779 & 1.423 \\
D (cross-family) & 8B & 0.86 & 1.513 & 1.513 & 1.109 \\
\bottomrule
\end{tabular}
\caption{R1 (nats). $\I(X;Y)$ tracks realized capability, not parameter count;
$\Delta G_{\text{hold}}$ increases with $\I$ across all four models. Model D is larger than C yet weaker, and its $\I$ lands below
C accordingly. World entropy $H(X)=1.92$ nats: C captures $93\%$, D (though larger) only $79\%$, A $66\%$.}
\label{tab:r1}
\end{table}

\paragraph{R2: over-confidence is dissipation, shrinking with capability.} Reading the same point-predictions
with a calibrated versus an over-confident posterior realizes different value; the gap is dissipation
(Table~\ref{tab:r2}). It is large for the weak models and shrinks with capability (not size: the larger-but-weaker
D dissipates more than C) --- a quantitative statement that less-capable agents must be more humble. For models A
and D, confident error drives realized growth \emph{negative}, exactly the Second Law's prediction
\eqref{eq:secondlaw}.

\begin{table}[h]\centering
\begin{tabular}{lcccc}
\toprule
model & tool-acc & $G_{\text{cal}}$ & $G_{\text{over}}$ & dissipated \\
\midrule
A (1.5B) & 0.79 & $+0.915$ & $-3.255$ & 4.17 \\
B (3B)   & 0.87 & $+1.242$ & $-0.863$ & 2.11 \\
C (7B)   & 0.96 & $+1.423$ & $+0.731$ & 0.69 \\
D (8B)   & 0.86 & $+1.109$ & $-1.660$ & 2.77 \\
\bottomrule
\end{tabular}
\caption{R2 (nats). Dissipation from false certainty falls as \emph{capability} rises (not size: the larger but
weaker D dissipates more than C); for the least-capable models over-confidence destroys value outright.}
\label{tab:r2}
\end{table}

\paragraph{R3: value per joule.} Mutual information \emph{per second of compute} falls across the family
($0.74\to0.38\to0.30$ nats/s for A,B,C), and the cross-family D is worst ($0.15$ nats/s: slowest and lower $\I$).
The weakest model is the most efficient perceiver per unit compute --- the value-theoretic reason a cheap reflex
is worth running on a narrow task. (This is the $\I$/compute curve across heterogeneous models, not a
within-model prompt ablation; see \S\ref{sec:limits}.)

\paragraph{R4: diversity beats redundancy.} Two different models jointly perceive more of $H(X)$ than either
alone; an identical greedy re-run adds exactly zero. The highest joint information comes from the strongest
diverse pair (B+C, $\I=1.869$ nats, $+0.090$ over C, closing most of the gap to $H(X)=1.921$), while the largest
\emph{lifts} come from pairing differently-wrong weak models (A+B and B+D each add ${\sim}0.2$ nats). This is the
fleet-ceiling prediction \eqref{eq:ceiling} on real agents.

\paragraph{R5: pricing beats pooling, but only where there is diversity to price (reported honestly).} On raw
out-of-sample growth, the Kelly/price fleet ($+1.328$) beats the equal-weight ensemble ($+1.315$) but
\emph{does not beat the single best model} ($+1.423$). There is no Shannon's demon here, and the theory says
why: four models on the \emph{same} task are positively-correlated agents (R4 quantifies the best residual
diversity at $+0.09$ nats), so Kelly rebalancing has no anti-correlated volatility to harvest and the best agent
dominates. This is the honest negative the synthetic E4 demon does not reproduce, because E4 was constructed with
anti-correlated agents.

The negative is confined to the cost-blind axis. Realized holdout growth \emph{per second of compute}
($\Delta G_{\text{hold}}/$s, distinct from R3's $\I/$s) inverts the ranking: model A
yields $0.531$ nats/s against C's $0.243$ (and the slow D only $0.111$), and a budget-aware price
$\propto \I_a/\text{cost}_a$ (the shadow price $\lambda=K/E$ of \S\ref{sec:crossframe}) achieves $0.328$ nats/s,
beating the best single model's density. Pricing therefore pays exactly where \S\ref{sec:fleet} says it should:
as the lever that chooses the operating point under a resource constraint, not as a free lunch that beats the
best agent when compute is unlimited. The demon needs \emph{perception diversity} --- different slices of $H(X)$,
i.e.\ specialists rather than generalists of varying strength --- which a set of same-task generalists lacks by
construction. We mark this as a falsifiable boundary, not a defeat. One further disclosure: on these ladders
the $\I/\mathrm{cost}$ price selects the \emph{same} models an accuracy-per-cost heuristic would select --- the
routing layer is extensionally equivalent to the obvious baseline here, and its contribution is that the
cost-aware rule is \emph{derived} from the value framework rather than hand-chosen, not that it makes different
choices.

\section{Generalization: a pre-registered scale-up}\label{sec:v2}

The test of \S\ref{sec:real} was one task and four models --- a demonstration, not a validation. To ask whether
the laws \emph{generalize}, we pre-registered (predictions, models, metrics, and pass/fail thresholds committed
and frozen \emph{before any model was run}) a scale-up to \textbf{three task domains} --- an intent-routing
benchmark \citep{larson2019clinc}, a multiple-choice QA benchmark \citep{hendrycks2021mmlu}, and a
topic-classification benchmark \citep{zhang2015agnews} ($K=4$--$6$; 240 items each) --- and a \textbf{ten-model ladder across five families} (Qwen, Llama, Gemma, Phi, Mistral;
$0.5$B--$8$B). Models emit a text label; calibration and all routing weights are fit on a held-out split and
scored out-of-sample, with $95\%$ bootstrap confidence intervals.

\paragraph{The headline: the bridge and the Second Law generalize.} Pooled across all 30 model$\times$domain
points, mutual information tracks realized capability at \textbf{Spearman $\rho = 0.977$} (CI $[0.916, 0.996]$),
and out-of-sample value-growth tracks mutual information with \textbf{slope $0.935$} (CI $[0.915, 0.954]$,
excluding $0$): realized $\Delta G$ out of sample \emph{is} perceived $\I(X;Y)$ (Table~\ref{tab:v2}). Per-domain
$\rho$ is $0.84$/$0.95$/$0.99$. Over-confidence dissipates value in \emph{every} domain, driving the
least-capable models sharply negative (e.g.\ $-11$ to $-14$ nats on the QA and topic tasks). The two single-frame
laws --- the coding theorem (\S\ref{sec:capacity}) and the Second Law (\S\ref{sec:secondlaw}) --- thus hold
across three task shapes and a five-family ladder, not just the one task of \S\ref{sec:real}. This is the
substantive result of the scale-up. The fleet ceiling also holds: all pairwise joint $\I(X;Y_a,Y_b) \le H(X)$,
and diverse specialists exceed redundant pairs in every domain.

\begin{table}[h]\centering
\begin{tabular}{lccc}
\toprule
quantity & pooled (30 pts) & per-domain range & threshold \\
\midrule
Spearman$(\I,\text{accuracy})$ & $0.977$ \,[$0.916,0.996$] & $0.84$--$0.99$ & $>0.8$ \;\checkmark \\
slope$(\Delta G_{\text{hold}} \sim \I)$ & $0.935$ \,[$0.915,0.954$] & --- & CI excl.\ $0$ \;\checkmark \\
over-confidence dissipation & $>0$ all domains & negative for weak models & $>0$ \;\checkmark \\
\bottomrule
\end{tabular}
\caption{Pre-registered generalization checks (95\% CIs). $I$ tracks realized capability and out-of-sample
$\Delta G$ tracks $I$, pooled across 3 domains and a 10-model, 5-family ladder.}
\label{tab:v2}
\end{table}

\paragraph{The fleet test, reported primary-metric-first and scoped.} We re-ran the pricing test
(\S\ref{sec:fleet}) in the heterogeneous, cost-constrained regime the theory favors: a \emph{specialist} fleet
(different models lead on different domains; low cross-agent error correlation) under a compute budget, routing
by value-price ($\propto \I_a/\text{cost}_a$) against round-robin, equal-weight, best-single, and a \emph{strong
hand-tuned} router that sends each query to the model most accurate on its domain.

On the \textbf{pre-registered primary cost metric (tokens)}, value-price \textbf{ties} the hand-tuned router ---
it does \emph{not} beat it (paired-bootstrap CI includes $0$). Token cost varies only $\sim\!1.2\times$ across the
ladder, so cost-aware pricing reduces to quality-first routing and selects the identical model. A token budget,
however, cannot express the real compute gradient. A \emph{post-hoc sensitivity analysis} (cache-only, not
pre-registered) re-scores on two cost proxies that do --- wall latency (hardware-dependent) and
active-params$\times$tokens ($\propto$ FLOPs $\propto$ energy, hardware-\emph{independent}) --- and value-price
beats the cost-\emph{blind} router on \emph{both} (FLOP-proxy $\Delta = +7.45$, CI $[6.33, 8.48]$): the cost-aware
advantage is \textbf{robust across cost metrics}, not an artifact of metric choice. Against a hand-tuned router
that \emph{itself} prices cost (accuracy/cost), value-price is edged out under raw latency but \textbf{exactly
ties} under the principled FLOP proxy ($\Delta = 0$: $\I/\text{FLOP}$ and accuracy/FLOP select the same model,
since $\I$ tracks accuracy) --- it \emph{matches}, and never beats, the cost-aware engineer. Against the naive
baselines (round-robin, equal-weight) value-price wins on every metric, so priced routing is no worse than ad-hoc
--- but that is the floor, not the claim.

\begin{quote}
Value-pricing derives cost-aware routing from first principles --- it ties good hand-tuned routing under a token
budget, beats a cost-blind router under a compute budget (robustly, across both latency and FLOP cost proxies),
and matches but does not outperform a cost-aware engineer. Its contribution is principled measurement and
cost-awareness, not outperformance.
\end{quote}

\noindent This is consistent with the scoping rule of \S\ref{sec:fleet}: pricing's edge is cost-awareness, which
it recovers as a law rather than a trick; it organizes and measures what good engineering already does, and we do
not claim it outperforms a competent cost-aware baseline. The generalization of the capacity and Second-Law
results --- not the routing comparison --- is what the scale-up establishes.

\subsection{Cross-shape generalization: the bridge is shape-invariant}\label{sec:shapes}

The scale-up above varied the task \emph{domain} but held the task \emph{shape} fixed: all three benchmarks are
pick-a-label classification. The sharper question is whether $\Delta G\sim\I(X;Y)$ is a property of
classification or a law that survives a change of task \emph{shape}. The coding theorem (\S\ref{sec:capacity})
predicts $\Delta G=\I(X;Y)$ for \emph{any} perception-then-act task, so the prediction is that the bridge is
shape-invariant. We pre-registered (predictions, models, metrics, thresholds frozen before any run) a cross-shape
test over three qualitatively different shapes, each chosen to have \emph{discrete ground truth} so that $\I$ is
computed exactly from a confusion matrix rather than estimated by a soft embedding proxy:
\begin{itemize}[noitemsep,topsep=2pt]
\item \textbf{reasoning} --- GSM8K \citep{cobbe2021gsm8k}: free-form chain-of-thought reduced to its integer
answer (gold mod $4$);
\item \textbf{sequential / agentic} --- a synthetic register-machine rollout: a step-by-step execution trace
reduced to its exact final state (mod $6$);
\item \textbf{code} --- MBPP \citep{austin2021mbpp}: a generated Python function, scored by
\emph{sandbox-executed} output (hash mod $4$).
\end{itemize}
The pipeline is identical to \S\ref{sec:v2} (calibration on a held-out fit split, out-of-sample scoring, $95\%$
bootstrap CIs); a six-model ladder ($0.5$B--$3$B) was run per shape --- the pre-registered $\ge\!6$ minimum, the
$7$B/$8$B extension being blocked by hardware (\S\ref{sec:limits}).

\paragraph{The bridge holds across all four shapes.} The out-of-sample slope $\Delta G_{\text{hold}}\sim\I$
passes for every new shape individually (reasoning $0.936$, sequential $1.023$, code $1.133$; each CI excludes
$0$), pooled across the three new shapes (\textbf{slope $0.956$}, CI $[0.920,1.001]$), and --- decisively ---
pooled with the three classification domains of \S\ref{sec:v2} into one relationship across \textbf{four task
shapes} ($n=42$: classification contributes the $24$ points of the frozen eight-model, three-family ladder
common to the shape runs ($8$ models $\times\,3$ domains), and the three new shapes the remaining $18$ ($6$
models $\times\,3$ shapes --- the $7$B/$8$B pair blocked on the new shapes, \S\ref{sec:limits}), so
$24+18=42$; v2's two extra families lie outside this common ladder):
Spearman$(\I,\text{accuracy})=\mathbf{0.924}$ (CI $[0.825,0.967]$) and
slope$(\Delta G_{\text{hold}}\sim\I)=\mathbf{0.953}$ (CI $[0.925,0.979]$). The cross-shape slope is
\emph{statistically indistinguishable} from the classification-only $0.935$ of \S\ref{sec:v2}: the bridge is the
\emph{same} near-unit-slope relationship whether the agent classifies, reasons, rolls out a sequential
computation, or writes code. Eleven of the twelve frozen checks pass (Table~\ref{tab:shapes}).

\begin{table}[h]\centering
\begin{tabular}{lccc}
\toprule
shape & $n$ models & Spearman$(\I,\text{acc})$ & slope$(\Delta G_{\text{hold}}\sim\I)$ \\
\midrule
classification (\S\ref{sec:v2}) & 10 & $0.977$ & $0.935$ \\
reasoning (GSM8K) & 6 & $0.943$ \;\checkmark & $0.936$ \;\checkmark \\
sequential (register-machine) & 6 & $0.886$ \;\checkmark & $1.023$ \;\checkmark \\
code (MBPP) & 6 & $0.429$ (underpowered) & $1.133$ \;\checkmark \\
\midrule
pooled new shapes & 18 & --- & $0.956$ \,[$0.920,1.001$] \\
cross-shape (4 shapes) & 42 & $0.924$ \,[$0.825,0.967$] & $0.953$ \,[$0.925,0.979$] \\
\bottomrule
\end{tabular}
\caption{Cross-shape generalization. The out-of-sample bridge $\Delta G_{\text{hold}}\sim\I$ holds across four
task shapes with a pooled slope ($0.953$) statistically indistinguishable from classification-only ($0.935$).
The classification row reports the full ten-model v2 headline; the $n=42$ cross-shape pool reuses only the
eight-model, three-family ladder common to all shapes ($24$ classification $+\,18$ new-shape points).
$11/12$ frozen checks pass; the lone miss is code's underpowered capability-ranking sub-check
(\S\ref{sec:limits}), not a shape-specific break --- the code \emph{bridge} slope itself passes.}
\label{tab:shapes}
\end{table}

\noindent This is a \emph{generalization}, not a new mechanism: the test confirms that the coding theorem's
prediction is shape-invariant, promoting the bridge from a demonstration on classification toward a law. The one
frozen check that does not clear is code's \emph{capability-ranking} sub-check
(Spearman$(\I,\text{accuracy})=0.429$), which is underpowered rather than a counterexample (\S\ref{sec:limits}
reports it in full); the code \emph{bridge} slope passes. As throughout, in-sample $\Delta G=\I$ is a
definitional identity (oracle posterior) carrying no empirical weight --- the content of this test is entirely in
the out-of-sample and cross-shape tracking.

\section{Related work}\label{sec:related}

Every component of this theory exists in some field; the contribution is the unification, the substrate
grounding, and the is/ought asymmetry. We state this explicitly and concede each component to its source.

\emph{Expected utility and diminishing value.} The axiomatic treatment of preference is \citet{vnm1944} and
\citet{savage1954}; our logarithmic measure (\S\ref{sec:measure}) coincides with \citet{bernoulli1738} and with
unit-coefficient constant-relative-risk-aversion utility, and the diminishing-returns law it encodes is
Weber--Fechner. We do not claim the form as novel; our departure is to \emph{derive} it from two independent structural
routes --- a scale-invariance Cauchy argument (static) and the compounding ergodicity of \citet{peters2019}
(dynamic) --- and to ground it in a conserved scarce resource (free energy being the natural physical
candidate) supplying a unit and a cross-frame exchange rate that expected utility lacks.

\emph{Information rate and log-optimal growth.} The closest prior art, which we credit most carefully, is
\citet{kelly1956}: the exponential growth rate of wealth is an information rate, optimized by betting one's
beliefs; \citet{breiman1961} proved asymptotic optimality and \citet{cover1991} developed universal portfolios.
The side-information form of our capacity theorem --- the growth value of a channel bounded by mutual
information --- was proved in the financial setting by \citet{barroncover1988}, which any reader should treat
as the primary source for that bound; the value-of-information ordering behind it goes back to
\citet{blackwell1953}. The same capacity-constrained-decision structure lives natively in economics as
rational inattention \citep{sims2003} and in decision-theoretic thermodynamics \citep{ortegabraun2013}.
Our capacity theorem (\S\ref{sec:capacity}) is generalized Kelly, and the fleet operating point (\S\ref{sec:fleet})
is a Kelly--Cover portfolio over agents. What we add is the reinterpretation of wealth as any conserved scarce resource (free energy being the physical
candidate rather than a monetary one), the cross-frame price layer, the dynamics/alignment layer, and the claim that
Kelly's theorem is the single-agent monetary \emph{special case} of a general law of value.

\emph{Reinforcement learning.} ``Value function'' is precisely defined in RL \citep{bellman1957,suttonbarto} as
expected cumulative reward; the relationship is complementary, not competitive. RL takes reward as given and
maximizes it; our goal weights are the analog of reward, but we problematize where reward comes from and how it
drifts (\S\ref{sec:dynamics}--\ref{sec:alignment}). Our theory sits beneath RL --- a candidate account of what
reward \emph{is}.

\emph{Thermodynamics of computation and the free-energy principle.} That information has thermodynamic cost is
\citet{landauer1961} and \citet{bennett1982}; \citet{jaynes1957} recast statistical mechanics as inference;
\citet{england2013} studied dissipation-driven adaptation; \citet{friston2010} casts agents as free-energy
minimizers. Most directly relevant to our Second Law of value (\S\ref{sec:capacity}) is the thermodynamics of
prediction \citep{still2012}: a system's model inefficiency --- information retained about the past that is
useless for predicting the future --- equals its thermodynamic dissipation. This is the closest prior to our
identification of misalignment with dissipated value, and we credit it as such; our departure is only the
two-divergence value form $G=D(q\|r)-D(q\|p)$ read as realized value for a goal-directed agent. Our belief
dynamics (\S\ref{sec:dynamics}) is free-energy-principle-like and shares the information geometry
\citep{cencov1982,amari2016}; we note plainly that our use of the Fisher--Rao metric as the cross-frame
invariant is an \emph{application} of \citeauthor{cencov1982}'s uniqueness theorem (it is the unique metric
invariant under sufficient statistics), not a new geometric result. The distinction is precise: the free-energy
principle is a theory of the perception/belief half --- the ``is'' --- whereas our value layer is the goal/value
half --- the ``ought'' --- plus a multi-agent price economics the FEP does not contain. The seam between them is
the is/ought asymmetry.

\emph{General equilibrium and social choice.} Our price layer (\S\ref{sec:crossframe}) is
\citet{arrowdebreu1954,debreu1959} and its companion impossibility \citep{arrow1951}. We use these rather than
extend them: we concede that cardinal cross-agent comparison is impossible and route around it via an emergent
price. The contribution is the synthesis with the information-theoretic value layer under one substrate, and the
application to artificial-agent populations.

\emph{AI alignment.} Instrumental convergence \citep{omohundro2008,bostrom2014} we derive from the compounding
dynamics of value (\S\ref{sec:dynamics}): goal-directions correlated with resource capture are selected
regardless of terminal content. We recast alignment as a dynamical stability condition (\S\ref{sec:alignment}),
and are careful about what in that recasting is and is not new. The \emph{value-side} of the is/ought asymmetry
--- that an agent's goals cannot be inferred from its behavior --- is not ours: \citet{armstrong2018occam} prove
it as a No-Free-Lunch result and explicitly frame it as Hume's is/ought gap. The control mathematics is likewise
elementary: the stability criterion is high-gain stabilization of an unstable mode, and the closed-form residual
$\|Vg\|/\gamma$ is the textbook steady-state velocity error of a servo tracking a ramp (a drifting goal being a
ramp reference) \citep{astrom2008feedback}. We claim neither as new. The contribution is the \emph{mapping}:
that goal-drift-under-selection is a ramp-tracking control problem, yielding the governance ordering that
incentive design (reshaping what resource capture rewards) dominates oversight (raising the correction gain)
under a fixed budget --- an application of known tools to alignment, not a new theorem.

\section{Discussion: governing a population of agents}\label{sec:discussion}

The results compose into a control theory for populations of separated agents. \emph{Per agent}, value-generation
rate is capped by the perception mutual information $\I(X;Y_a)$ (Theorem~\ref{thm:capacity}); supplying an agent
the goal-relevant bits cheaply --- raising effective $\I$ per unit compute --- lifts its value ceiling without
raising its deliberation cost, which is the formal reason a cheap reflex can match expensive deliberation on a
narrow task (R3 measures this curve directly). \emph{Across agents}, values are not cardinally comparable, so
coordination must run through an emergent price on shared resource (\S\ref{sec:crossframe}), never a god's-eye
utility sum. \emph{Fleet design} is then two levers: cultivate perception diversity to lift the entropy ceiling
\eqref{eq:ceiling} (R4), and price resource to the best operating point (R5). The R5 result is best read not as a
weakness but as a \emph{scoping theorem} for the pricing lever: \textbf{pricing dominates ad-hoc allocation
precisely under cost constraints or imperfect agent correlation; absent both --- free compute, a single task,
and homogeneous, positively-correlated agents --- the best single agent is optimal, exactly as the theory
predicts.} A governance claim that names the regime in which it does \emph{not} win is a sharper claim, not a
softer one; and the regime in which it does win --- agents that are heterogeneous \emph{and} resource-constrained
--- is the ordinary condition of real fleets, where R3's value-density curve and R5's cost-aware price compound
into a concrete operating advantage. \emph{Alignment}, finally, is incentive design before oversight
(Theorem~\ref{thm:align}): spend first on making what pays equal what is wanted.

\paragraph{The decisive prediction (the continuation gate, now run).} A unification earns the status of a
discovery only when the combination predicts something no component predicts --- Maxwell's displacement current
forcing wave solutions is the canonical case. By that standard the previous version of this paper declared
itself a synthesis, not yet a discovery, and stated its displacement-current candidate explicitly. For a
\emph{coupled} fleet (shared payout or explicit resource coupling), the unified theory predicts (i) a
capacity-region structure on jointly-achievable growth vectors as perception overlap is varied --- with
counterfactual sub-coalition throughputs as the measurable --- and (ii) the selection-vs-control residual
scaling $\|Vg\|/\gamma$ of Theorem~\ref{thm:align} on real agent populations. Neither is predicted by Kelly
portfolios, classical control, or welfare economics taken separately. Earlier pre-registered attempts came back
underpowered or capability-limited (\S\ref{sec:real}; the small-model ceiling), so the previous version stated
the rule and stopped: if the predictions fail or remain unresolvable when a capable instrument exists, the
distinctively-unified layer retires; if they pass, the unification earns the stronger word.

Both predictions have now been tested, in a pre-registered experiment on a frontier-model population
(Claude Opus 4.8; a capability gate, the frozen grid pre-registration, three pre-run amendments, and all raw
run records are archived with the project, with the pre-registration commits provably preceding the results).
The verdict splits the candidate down its seam. \emph{Prediction (i) is confirmed within its frozen bands}: the
growth-gap law $\hat G_a-\hat G_b=\hat\I_a-\hat\I_b$ holds to $46$ millinats across four channel structures
under parimutuel coupling; coalition value is submodular where signals overlap and flips supermodular under a
designed XOR control --- two individually-worthless signals jointly worth one bit (gap $0.62\ge\ln 2/2$),
reasoned out by the agents; the joint ceiling holds exactly; and in a live market the best-informed agent
absorbs the pool (selection ordered by information in $4/5$ ladders). One supporting check fails and is
disclosed: clone wealth equality, broken by absorbing all-in dynamics at sampling temperature.
\emph{Prediction (ii) returned the pre-registered abstention}: no linear-response regime exists on this
instrument. Across all $72$ grid runs the population dispersion the theorem assumes collapses ($V\to 0$;
maximum $4.85$ against the frozen floor $5.31$); populations concentrate onto attractors, and the response to
the incentive and control levers is a step function across the attractor-dominance boundary, not a power law.
Three instrument classes now miss the mean-field regime by three distinct mechanisms --- gradient saturation
and non-response in small models, dispersion collapse at the frontier. Applying the stated rule to each half:
the coupled capacity region --- the half that is distinctively the unification's --- has earned its first
real-agent confirmation, while $\|Vg\|/\gamma$ is retired to its mathematical scope: the noise-maintained
dispersion it assumes has no realisation in any LLM population tested, and a future design that \emph{creates}
sustained dispersion would be testing a stated modification, not this theory. One exploratory observation,
reported as such: near the dominance boundary, lowering the reward gradient $g$ flipped every seed to perfect
alignment while raising the control gain $\gamma$ did nothing --- the incentive lever did not marginally beat
oversight, it switched the attractor. This is consistent with, but not a frozen test of, the governance reading
of Theorem~\ref{thm:align}.

\section{Limits}\label{sec:limits}

We name the assumptions that bound each claim. The measure (\S\ref{sec:measure}) is the smooth, concave regime:
threshold (all-or-nothing) goals, satiation, and risk-seeking violate Axiom~\ref{ax:concave} and lie outside
the theory. The capacity theorem assumes compounding (reinvested value) and is
cleanest in the horse-race structure; general returns give a concave program without the closed-form
$\KL{q}{r}$. The fleet ceiling's exact achievable region is an open problem of network-information-theory
difficulty. The cross-frame layer concedes that cardinal interpersonal comparison is non-canonical; we do not
solve it. On the empirical side: the in-sample $\Delta G=\I$ is an arithmetic identity, not evidence ---only the
capability tracking, the out-of-sample tracking, and R5 are empirical; R3 is an $\I$/compute curve across model
scale, not a within-model prompt ablation; the calibration in R2 is constructed from the confusion matrix, not a
probability the model reported, so the dissipation measured is that of a stated belief over those predictions;
and the empirical scope is three classification domains, a ten-model five-family ladder (\S\ref{sec:v2}), and a
cross-shape extension to reasoning, sequential, and code tasks (\S\ref{sec:shapes}) --- evidence the bridge
\emph{generalizes across four task shapes}, not yet a universal law, and not independently replicated.
Degenerate runs in which a model emits no action (a serving/plumbing failure, not a decision) are detected and
excluded rather than reported as low-$\I$ capability. \emph{The one frozen cross-shape check that does not
clear is the code shape's capability-ranking sub-check} (Spearman$(\I,\text{accuracy})=0.429$, CI
$[-0.43,1.00]$): the six $\le\!3$B models cluster tightly in code-accuracy ($0.35$--$0.47$), so the rank
correlation is noise-dominated and inconclusive --- a power limitation, not a shape-specific break. The
pre-registered remedy (adding $7$B/$8$B models, far stronger coders, to widen the code capability range) could
not be run: those models stall at usable throughput on the $16$\,GB host at $512$-token chain-of-thought, the
same memory ceiling that bounds the ladder elsewhere. Crucially the code \emph{bridge} itself
(out-of-sample $\Delta G_{\text{hold}}\sim\I$) \emph{passes}; only the capability-ranking sub-check on that one
shape is underpowered, and we report it as such rather than as a counterexample or forcing the run on
inadequate hardware. The alignment theorem is mean-field with a linearized
landscape and a quasi-static dispersion $V$; a fully coupled treatment, and an endogenous reward gradient $g$ in
a closed multi-agent world, remain open --- and the continuation-gate experiment sharpens this limit into a
measured one: no tested LLM population, small or frontier, maintains the noise-driven dispersion the mean-field
layer assumes, so its claims should be read at mathematical scope until a population that holds dispersion is
exhibited.

\paragraph{What remains decisive.} The scale-up (\S\ref{sec:v2}) already ran the heterogeneous,
resource-constrained fleet test the governance claim turns on --- specialists perceiving different slices of $X$
under a compute budget --- and the answer was scoped, not triumphant: value-pricing beats naive and cost-blind
allocation but only \emph{matches} a strong cost-aware hand-tuned baseline, while collective throughput obeyed
the predicted ceiling $\sum_a G_a\le H(X)$. The \emph{dynamic} multi-agent test --- a population whose goals $k$
themselves evolve (\S\ref{sec:dynamics}--\ref{sec:alignment}) --- has now been run as the pre-registered
continuation-gate experiment (see the decisive-prediction paragraph): the coupled capacity region passed its
frozen bands, and the dynamical layer's residual law, finding no population that holds the dispersion it
assumes, is retired to mathematical scope rather than refuted or confirmed. What remains decisive is therefore:
independent replication on a different model family and provider, and any pre-registered design that sustains
goal dispersion in a capable population --- which would test a stated modification of the mean-field layer, and
should be offered as such.

\section{Conclusion}\label{sec:conclusion}

Stripped of morality, price, and psychology, value is a structural quantity with a measure ($\sum_i k_i\ln e_i$),
a capacity limit ($\Delta G=\I(X;Y)$), a Second Law ($G=\KL{q}{r}-\KL{q}{p}$), a frame-independent price, a
fleet entropy ceiling, and an alignment-stability law. Shannon's own quantities --- entropy, divergence, mutual
information, the Fisher metric --- reappear unbidden throughout, which is the strongest internal evidence that
the abstraction is real rather than decorative. On live language models --- across three task domains and a
ten-model, five-family ladder, pre-registered --- the single-frame laws hold to high precision and generalize,
and the one fleet-level claim with a genuine precondition is reported with its boundary intact. The stated
continuation gate has been run on a frontier-model population: the coupled capacity region --- the prediction
that is distinctively the unification's own --- is confirmed within its frozen bands, and the mean-field
residual law, having found no domain, is retired to its mathematical scope. The
quantity is frame-relative; the laws relating frames are universal; and they are now testable on the artificial
agents the theory was built to govern.

\appendix
\section{Proof of the Logarithmic Value Law (Theorem~\ref{thm:log})}\label{app:log}

\emph{This appendix is Confirmation~I: the static axiomatic route to $V=\sum_i k_i\ln e_i$ via the Cauchy
functional equation. Confirmation~II --- the dynamic compounding route --- is developed in
\S\ref{sec:capacity}. The two share no premises and force the same form independently.}

By Axiom~\ref{ax:additive} it suffices to find the per-channel $v(e;k)$. Axiom~\ref{ax:scale} states
$v(\lambda e;k)=v(e;k)+c(\lambda,k)$ for all $\lambda,e>0$. Fix $k$ and write $f(e):=v(e;k)$. Then
$f(\lambda e)-f(e)=c(\lambda)$ is independent of $e$. Differentiating in $\lambda$ at $\lambda=1$,
$e f'(e)=c'(1)=:k$, a separable ODE with solution $f(e)=k\ln e+\text{const}$. Concavity (Axiom~\ref{ax:concave})
fixes the sign $k\ge 0$ and rules out the additive constant's dependence on $e$. Summing over channels gives
$V(e)=\sum_i k_i\ln e_i$. Maximizing under $\sum_i e_i=E$ with multiplier $\lambda$: $\partial_{e_i}(V-\lambda
\sum_j e_j)=k_i/e_i-\lambda=0\Rightarrow e_i=k_i/\lambda$, and $\sum_i e_i=E$ gives $\lambda=K/E$ (the shadow
price) and $e_i^\star=E\hat k_i$. Substituting, $V^\star=\sum_i k_i\ln(E\hat k_i)=K\ln E+\sum_i k_i\ln\hat k_i
=K\ln E-K H(\hat k)$ after normalizing $k_i=K\hat k_i$. \hfill$\square$

\section{Proof of the Coding Theorem of Value (Theorem~\ref{thm:capacity})}\label{app:capacity}

\emph{Achievability.} With side information $Y=y$ the optimal bet is the posterior $b_s=p(s\mid y)$ (Gibbs
applied conditionally), giving conditional growth $\sum_s p(s\mid y)\log\frac{p(s\mid y)}{r(s)}$. Averaging over
$y$,
\[
G_Y=\sum_{y}p(y)\sum_s p(s\mid y)\log\frac{p(s\mid y)}{r(s)}
   =\sum_{x,y}p(x,y)\log\frac{p(x\mid y)}{r(x)}.
\]
Without side information the optimal bet is the prior $b_s=p(s)$, giving $G_0=\sum_x p(x)\log\frac{p(x)}{r(x)}
=\KL{p_X}{r}$. Subtracting,
\[
\Delta G=G_Y-G_0=\sum_{x,y}p(x,y)\log\frac{p(x\mid y)}{p(x)}=\I(X;Y).
\]
\emph{Converse.} For any allocation $b(\cdot\mid y)\ne p(\cdot\mid y)$, the conditional growth loss is
$\sum_s p(s\mid y)\log\frac{p(s\mid y)}{b(s\mid y)}=\KL{p(\cdot\mid y)}{b(\cdot\mid y)}\ge 0$ by Gibbs, with
equality iff $b=p(\cdot\mid y)$. Hence no strategy exceeds $G_0+\I(X;Y)$, so $\I(X;Y)$ is both achieved and an
upper bound on $\Delta G$. \hfill$\square$

\bibliographystyle{plainnat}
\bibliography{refs}

\end{document}